# Timing decisions as the next frontier for collective intelligence

Albert B. Kao[1]*, Shoubhik Banerjee[1], Fritz Francisco[1], Andrew M. Berdahl[2]*

[1] Biology Department, University of Massachusetts Boston, Boston, MA, USA

[2] School of Aquatic and Fishery Sciences, University of Washington, Seattle, WA, USA

* Correspondence: albert.kao@umb.edu (A. B. Kao), berdahl@uw.edu (A. M. Berdahl)

## Highlights

- Group living is ubiquitous across environments and taxa. One important benefit of group living is an improved ability to make accurate decisions – collective intelligence.
- Research on collective intelligence has focused on spatial (where to go), as opposed to temporal (when to go), decisions and it is through this spatial lens that our intuition of collective intelligence has developed.
- We illustrate several key ways in which temporal decisions are fundamentally different from their spatial counterparts. These differences make it unclear whether or not the known collective intelligence mechanisms apply to temporal decisions, or if as-yet-unknown mechanisms are at play.
- We encourage explicit studies of collective intelligence in the time domain. This is especially crucial in the Anthropocene, where changes in the climate are causing shifts in timing decisions, with little-known consequences.

## Abstract

The past decade has witnessed a dramatically growing interest in collective intelligence – the phenomenon of groups having an ability to make more accurate decisions than isolated individuals. However, the vast majority of studies to date have focused, either explicitly or implicitly, on *spatial* decisions (*e.g.*, potential nest sites, food patches, or migration directions). We highlight the equally important, but severely understudied, realm of *temporal* collective decision-making, *i.e.*, decisions about when to perform an action. We argue that temporal collective decision making is likely to differ from spatial decision making in several crucial ways and probably involves different mechanisms, model predictions, and experimental outcomes.

We anticipate that research focused on temporal decisions should lead to a radically expanded understanding of the adaptiveness and constraints of living in groups.

## Collective intelligence in space and time

The idea that animals may make more accurate decisions as a group than individually – collective intelligence – is an intriguing hypothesis that has received a substantial amount of research attention. Theoretical models have uncovered a multitude of mechanisms by which decision accuracy can be improved collectively, including a simple averaging of individual errors (e.g., the "many wrongs principle") (Simons 2004, King and Cowlishaw 2007, Berdahl et al. 2018), context-dependent leadership or social learning (Couzin et al. 2005, Laland 2004), adjusting social network structure (Kao and Couzin 2019, Becker et al. 2017), and emergent sensing or collective learning (where the group can effectively achieve a task or learn about a feature of the environment that individuals cannot perceive) (Berdahl et al. 2018, Kao et al. 2014, Biro et al. 2016). Alongside this rapid progress in theoretical modeling, an increasing number of empirical studies has also revealed collective intelligence in real animals (e.g., Berdahl et al. 2013, Webster et al. 2017, Ward et al. 2011). Collective intelligence has also been a major focus in many human contexts (Kameda et al. 2022), including team performance (Woolley et al. 2010, Krause et al. 2010, Kurvers et al. 2015), medical diagnoses (Centola et al. 2023, Kurvers et al. 2023, Radcliffe et al. 2019, Wolf et al. 2015), and AI (Ha and Tang 2022).

Importantly, however, the bulk of this work has focused, either explicitly or implicitly, on spatial decisions. In other words, most models and experiments study situations in which animals in groups decide *where* to go. In some studies, the directness of a trajectory to the goal location is the primary measure of collective intelligence (e.g., Mueller et al. 2013, Sasaki and Biro 2017). In others, animals in groups must decide between discrete locations in space, such as food patches (e.g., Couzin et al. 2011, Lihoreau et al. 2016, Strandburg-Peshkin et al. 2013), nest sites (Seeley et al. 2012, Sasaki and Pratt 2011, Sasaki et al. 2013), or to avoid a potential predator (e.g., Ward et al. 2011, Papadopoulou et al. 2022). Many theoretical models are abstract and simply consider a number of discrete options (e.g., Kao and Couzin 2014, Lee and Lucas 2014, Churchland et al. 2008, Yang et al. 2021, Sridhar et al. 2021) or a continuous space of options (Axelrod et al. 2021, Kameda et al. 2022) without explicit reference to space. Nonetheless, these too can be mapped onto spatial decision tasks – discrete locations in space and directions of travel, respectively. Therefore, much of our current intuition of collective intelligence in animal groups has effectively been derived from studies of spatial decision making.

By contrast, temporal decision making – *when* to perform an action – has been much less studied in a collective intelligence context (Helm et al. 2006, Oestreich et al. 2022). And yet, timing decisions are just as consequential to the fitness of animals (Gienapp and Bregnballe 2012, Scheuerell et al. 2009, Bontekoe et al. 2023). For example, migration is a ubiquitous phenomenon exhibited by animals across many clades, and deciding when to migrate can have major fitness consequences by affecting the fat stores that an animal will have when beginning its journey, the weather conditions encountered along the way, the efficiency of the migration route, and the breeding sites and mates available when it arrives at its destination (Gienapp and Bregnballe 2012, La Sorte et al. 2015, Bauer et al. 2016, Muller et al. 2018, Sergio et al. 2014, Flack et al. 2016). On shorter time scales, animal groups decide when to move from one location to another throughout the day (Stewart and Harcourt 1994, Black 1988, Strandburg-Peshkin et al. 2015). This can strongly affect, for example, the energy intake rate that an individual experiences (Davis et al. 2022). Furthermore, animals must decide when to flee from an encroaching potential predator (often quantified in the literature as a “flight initiation *distance*” rather than a *time*) or more generally when to react to environmental cues (Warkentin 2011, Majoris et al. 2022), which can starkly determine whether the individual lives or dies (Cooper and Frederick 2007, Tatte et al. 2018, Klamser and Romanczuk 2021).

Such timing decisions are often made by social animals living in groups. Indeed, synchronizing timing decisions across group members is essential to maintaining group cohesion, which in turn is often crucial to accruing the benefits of sociality (Conradt and Roper 2010, Krause and Ruxton 2002). As with spatial decisions, animals experience uncertainty about the optimal time to perform an action. Therefore, collective timing decisions may, in addition to maintaining cohesion, have the potential to improve the precision and accuracy of the timing of different actions by pooling the noisy estimates of multiple group members, or other mechanisms. However, heterogeneity within the group may also cause different group members to have intrinsically different optimal leaving times, which can cause conflict and “consensus costs” within the group (Conradt and Roper 2003, Conradt and Roper 2005).

Timing decisions have been studied extensively, but not in the context of collective intelligence. This includes the timing of laying or hatching of eggs (e.g., Tomás 2015, Kluen et al. 2011), migration (Gienapp and Bregnballe 2012), flight initiation distance (Cooper and Frederick 2007), and intertemporal decision-making (related to time discounting) (Namboodiri et al. 2014). Accordingly, theory has been developed to predict optimal leaving times through the marginal value theorem (Charnov 1976), optimal stopping theory (Freeman 1983, Shiryaev 2007), and in changing environmental conditions (McNamara et al. 2011, Winkler et al. 2014). Among social species, research has demonstrated that social influence can affect the timing decisions of animals across a wide variety of taxa (Oestreich et al. 2022), including birds (Helm et al. 2006, Dibnah et al. 2022), mammals (Gall et al. 2017) and fish (Berdahl et al. 2017). Some theory has been developed to examine how consensus costs can lead to different decision mechanisms to evolve in timing decisions (Conradt and Roper 2003, Conradt and Roper 2005). And yet, to our knowledge, there has been no work explicitly examining collective intelligence in timing decisions.

This near exclusive focus on spatial, but not temporal, decisions leaves us with only a partial understanding of collective intelligence in animal groups. Here, we argue that temporal collective decisions are not only a crucial contributor to the fitness of social animals, but that the mechanisms and dynamics underlying collective temporal decision making are likely to be substantially different compared to spatial decision making. Therefore, developing new theory and experiments specifically to understand collective temporal decisions is a fertile domain that will yield new insights into the function, adaptiveness, and evolution of collective intelligence in animals.

## Some key differences between temporal and spatial collective decisions

Deciding among locations or directions in space, and among moments in time, differ in several important ways, which will almost certainly render our intuition gained from studies of the former inaccurate in making predictions about the latter. Here we detail some of these crucial differences but anticipate that others will likely be revealed with more research on collective temporal decision-making.

1. *Sequential ordering of time*

In spatial decision-making, multiple options are often available simultaneously, and animals can often freely sample these options in any order (Figure 1A). This can include discrete food patches (e.g., Sridhar et al. 2021, Sasaki et al. 2013, Seeley and Visscher 2004), distinct routes (Nagy et al. 2020, Sawyer et al. 2019), or continuous directions of movement (Chapman et al. 2011, Biro et al. 2007). An animal may then choose a direction or location that it has previously sampled, or one that has not yet been sampled. Such spatial decision making has been modeled using a wide variety of methods, including multi-armed bandit problems (Morimoto 2019), drift-diffusion models (Ratcliff et al. 2016), and hidden Markov models (Ylitalo et al. 2021).

By contrast, time can only proceed in a single direction. Temporal options are strictly ordered (*i.e.*, in time), and only one option is available at any given moment (*i.e.*, the present moment).

When a moment passes, it is forever lost to the animal as an option. Therefore, there is a choice asymmetry, with all moments in the past inaccessible to an animal, and only moments in the future (or the present) potentially accessible (Figure 1B). There is additionally an informational asymmetry: animals may sample moments only in the past (by experiencing them directly and retrieving them from memory) but cannot directly sample moments in the future; they can only make predictions about the future (Redshaw and Bulley 2018, Suddendorf and Corballis 2010).

For example, many animals must choose when to begin their seasonal migration. Their decision-making process can be mapped into a series of yes-no (*i.e.*, binary) decisions, whereby they decide whether or not to begin their migration at each passing moment (Oestreich et al. 2023). To aid in their decision, they may reference the weather conditions in previous days, their current physiological condition, and the decisions made by conspecifics and/or heterospecifics to extrapolate into the future to predict when the optimal migration time might be. However, they are not able to directly gain information about the future before it occurs; therefore, they can never be certain that their chosen migration time is indeed the optimal one when they make it.

It may be possible for animals in groups to harness collective intelligence to make more accurate timing decisions, but the decision mechanisms are likely to be qualitatively different compared to spatial decisions. Animals often signal when they want to depart but do not possess an explicit signal that they do not want to depart, leading to an asymmetry in the communication of preferences. A signal likely communicates that the animal is ready to leave *now* rather than specifically communicating a desire to leave at a particular time in the future. Finally, many signals are likely to be binary in nature, indicating a readiness to leave (or not leave) (Conradt and Roper 2003; but see Black 1988).

One potential conundrum that this poses is choosing when to begin signaling to go. If an individual begins signaling only when its perceived optimal leaving time has been reached, the group will almost certainly leave sometime *after* the individual's optimal time. The optimal time to begin signaling depends on the cost of leaving at a time other than the optimal one, as well as the signaling strategies of other group members. However, if all individuals in the group signal to leave early, then this may bias the collective leaving time to be earlier than is optimal for the individuals. These fundamental differences between spatial and temporal decision making necessitate new classes of decision models that can facilitate collective intelligence specifically in the time domain.

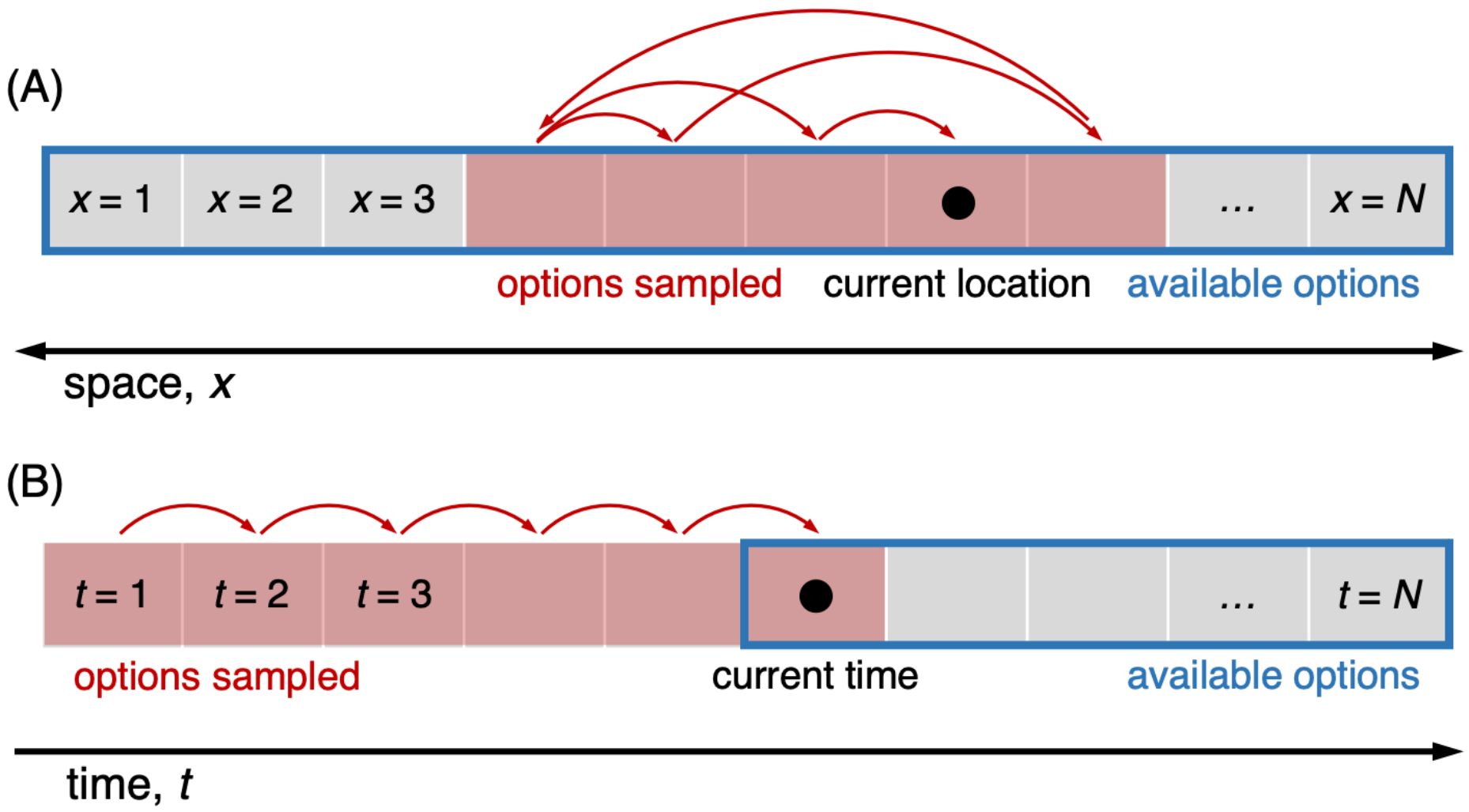

**Figure 1. Key differences between spatial and temporal decisions. (A)** For spatial decisions, many options can be sampled in a variety of orders and, in principle, all options remain available to the individual or group. **(B)** For temporal decisions, options are sampled strictly in order and once in the past, sampled options are no longer available. A hallmark, and challenge, of temporal decisions is the lack of overlap between sampled and available options. By contrast, for spatial decisions there is a large (potentially complete) overlap in sampled and available options. Both panels depict a discrete space/time option set (with arbitrary length/time units) but the same principles apply for continuous space/time.

2. *Asymmetric costs of too-early and too-late errors*

Empirically, there is often an asymmetry between the cost of performing an action too early and performing it too late (Conradt and Roper 2003 and stated in a more general context in Parker and Maynard Smith 1990). For example, when an animal is foraging and is approached by a potential predator, the animal must decide when to flee. Escape too early and it misses out on additional foraging opportunities (a relatively minor cost), but attempt too late and it will be eaten by the predator (a very high cost) (Cooper and Frederick 2007). Similarly, animals that forage in tidally-flooded river estuaries, such as bottlenose dolphins (Fury and Harrison 2011), suffer an opportunity cost if they return to the sea too early (relative to the onset of low tide) but risk being stranded and dying if they leave too late. For seasonally migrating animals, arriving late to their breeding grounds may mean settling for a worse nest site and missing out on mating opportunities, but arriving too early may mean deadly weather conditions (Møller 1994).

This asymmetry of costs is not built into most "wisdom of crowds" models where individuals in groups decide among a continuous set of values (*e.g.*, when to go, what direction to travel, numerosity estimation) (Conradt and Roper 2003, Conradt and Roper 2007, Conradt and Roper 2010). Indeed, in many of these models, individuals typically have a noisy estimate of the optimal value and adopt the average opinion as the collective decision usually results in improved decision accuracy (Golub and Jackson 2010, Kao et al. 2018).

By contrast, a simple average of estimates of the optimal value will not maximize fitness when an asymmetry of costs is present. Box 1 presents a simple model demonstrating that, in general, animals should learn to estimate a time on the shallower side of the fitness curve (earlier in time in the example in Box 1). However, the wisdom of crowds should allow a larger group of animals to better estimate a particular time, so that it can afford to perch closer to the objective optimal time, while smaller groups should choose a time further away from the optimal. If a social species exhibits fission-fusion dynamics, whereby the size of an animal's group can change frequently, these different group-size-specific optimal times can become problematic if an animal cannot accurately estimate the size of its group (Kao et al. 2014). In particular, learning the group-size-specific optimal time in a large group and then moving to a small group is likely to have larger fitness costs than the converse. In these situations, animals may simply learn the group-size-specific optimal time for a small group, thereby foregoing the potential benefits of the wisdom of crowds.

3. *Speed-accuracy trade-off is non-monotonic*

In decision theory, the "speed-accuracy trade-off" is a fundamental principle, where decision accuracy can be improved if the individual (or the group) takes more time to accumulate more information, or faster decisions can be made but at the cost of lower accuracy (Chittka et al. 2009, Bogacz et al. 2010). By contrast, the speed-accuracy tradeoff fails in the context of timing decisions. This is because the time spent gathering more information simultaneously decreases the set of temporal options available to an animal (Figure 1B). Specifically, waiting longer to make a decision may allow an animal to better estimate the optimal time to perform an action, but, the longer the animal waits, the more likely it is that this optimal time lies in the past (which is no longer accessible to the animal as an option due to the irreversible nature of time). Therefore, waiting longer may improve decisions at short time scales but will lead to poor decisions at long time scales, resulting in an optimal amount of time to gather information (Box 2). What decision strategies may be beneficial in these scenarios, and moreover, how collective intelligence may interact with these dynamics, is not known.

4. *Density-dependent strategies*

Game-theoretic considerations are often at play in timing decisions that may be absent in spatial decisions, due to the inherently directional nature of time. For example, there can be "finder's fees," whereby early deciders accrue more benefits than later ones by gaining access to more of a resource. Alternatively, an individual's fitness may be affected both by its absolute timing as well as relative timing compared to conspecifics (Iwasa & Levin 1995). In other words, leaving when other groupmates leave may be just as important as leaving at a particular time, in order to gain the benefits of group living (Krause and Ruxton 2002). How animals in a group negotiate different opinions among themselves in order to reach a consensus decision about a timing event is poorly understood (but see Conradt and Roper 2003, Conradt and Roper 2007, Conradt and Roper 2010).

These four examples illustrate that timing decisions and spatial decisions fundamentally differ from each other in crucial aspects, especially in a collective context. If we simply apply our intuition gained from the many studies of collective spatial decision making, then we are likely to make incorrect predictions about the behavior of animals making timing decisions. Therefore, new theoretical models are sorely needed that specifically describe the scenario of making decisions about when to perform an action. In particular, collective intelligence can arise during spatial decisions via several different known mechanisms, but it is not known which of these mechanisms apply to timing decisions, or if other, as-yet-undescribed mechanisms operate during timing decisions that allow for the emergence of collective intelligence.

## Collective timing in the Anthropocene

A rigorous understanding of collective timing decisions is especially important now that our planet is in the midst of substantial shifts in its climate. These changes may alter both the optimal time to perform an action, as well as the timing of environmental cues to which many species have evolved to respond (Cohen et al. 2018; Horton et al. 2020). Understanding the mechanisms through which collective intelligence can arise in timing decisions will also shed light on how robust these mechanisms may be to perturbations to the environment, such as gradual or sudden shifts in the timing of cues (Winkler et al. 2014, Shipley et al. 2020, Garcia et al. 2014, Crick and Sparks 1999, Millán et al. 2021, McCarty 2001). In addition to shifts in the mean time of events, climate change is also expected to increase the variability in timing (Garcia et al. 2014), which may alter the optimal amount of time a group should invest in collecting information for a timing decision (Fig. B2). Animals may also shift the timing of behaviors even when other environmental variables other than timing change (Millán et al. 2021).

Because of this, animals may need to increasingly rely on using information across more sensory modalities to estimate the best time to perform an action (*e.g.*, when to start a migration). With more uncertainty in the optimal timing of events, collective intelligence may also become an increasingly important method that social animals must use to reduce noise and make sufficiently accurate timing decisions to survive in the Anthropocene. However, without a theoretical understanding of the mechanisms at play in collective timing decisions, we currently do not know whether or not collective behavior may help, or harm, the fitness of social animals in the face of shifts in climate.

In parallel to the collective behavioral response of non-human animals to climate change are the policy decisions that humans may make, or fail to make, to stem the tide of climate change (Gerlagh et al. 2009, Ricke and Caldeira 2014, Hilbe et al. 2013, Domingos et al. 2020). Humans also face an asymmetry of costs here, where making policy changes too early may result in unnecessary economic costs, but making changes too late can lead to the global climate crossing several tipping points (IPCC 2022). Because of uncertainties in our climate models, it is probable that decisions should be made earlier than the objective optimal time.

## Concluding remarks

Collective behavior is extremely prevalent across scales of biological organization, from cells to social groups, and can lead to a variety of benefits to those organisms, including improved decision making (collective intelligence). Understanding what mechanisms can give rise to collective intelligence can help to explain the strategies and evolutionary drivers of social species as well as their broader effects on ecological scales (Westley et al. 2018). While collective intelligence among spatial options is increasingly well understood, there is nearly no work on collective intelligence among temporal options. Simple thought experiments reveal fundamental differences between space and time that are likely to render many of the well known “wisdom of crowds” mechanisms to be ineffective when making timing decisions. Focusing specifically on the time domain should provide fertile ground for new and influential contributions to the field of collective intelligence, thereby expanding our understanding of social animals and humans alike.

## Acknowledgements

A.B.K. acknowledges support from the NSF (BRC-BIO DBI-2233416). A.M.B. was supported by the H. Mason Keeler Endowed Professorship in Sports Fisheries Management.

## Outstanding questions

- When do simple ‘wisdom of crowds’ mechanisms (such as averaging) lead to better timing decisions, and when do they fail?
- When do emergent mechanisms (such as collective sensing or collective learning) lead to better timing decisions, and when do they fail?
- For both simple and emergent mechanisms for collective temporal decisions, how does accuracy scale with group size? Can we create models with testable predictions to guide empirical work?
- What new collective decision making mechanisms allow for collective intelligence in timing contexts but not spatial ones?
- How can we distinguish among collective decision making mechanisms that govern collective temporal decision making in lab or field experiments?
- How can we design experiments with a known optimal time to perform an action or identify this time in data from the field?
- What are the population-level or ecological implications (*e.g.*, Allee effects) of organisms making collective temporal decisions?
- How is temporal collective intelligence distinct from, and/or complementary to, consensus mechanisms used to simply maintain synchrony?
- What other differences between time and space make temporal decisions distinct from spatial decisions?

**Box 1. Model of asymmetric costs of too-early and too-late errors**

We illustrate through a simple mathematical model how the asymmetry of costs often observed in timing decisions confounds the prediction made by typical wisdom of crowds models (which is based on spatial decision-making). Consider a species of animal foraging in a tidally-flooded estuary. The amount of food that an individual consumes increases linearly with the amount of time spent in the estuary, but the risk of being stranded (and therefore death) increases exponentially the longer they remain. The overall fitness, as a function of time spent in the estuary, is described by $f = t - \exp(t)/b$ for this illustrative example, with $b$ = 20. The optimal time $t^*$ spent foraging in the estuary that maximizes fitness is $t^* = \log(b) \sim 3.00$ (Figure B1a, black curve).

However, the above calculation assumes that animals can measure time perfectly, which they generally cannot. Instead, we assume that an individual makes a noisy estimate of its desired leaving time, which is normally distributed with mean $t_D$ and standard deviation $\sigma = 2$.

We then consider groups ranging in size from $N$ = 1 to 31. Each individual makes a noisy estimate of its desired leaving time $t_D$ and signals to leave at that time. When half of the group has signaled, the group leaves. We scanned across $t_D$ from -1 to 3 and simulated the decision-making process 100,000 for each set of parameter values. We calculated the mean decision time and mean fitness for each combination of $N$ and $t_D$, and the value of $t_D$ that resulted in the highest fitness for that group size.

Larger groups are able to choose a desired leaving time $t_D$ very close to the objective optimal time because the wisdom of crowds allows that group to accurate identify that time (Figure B1a, purple), whereas smaller groups should opt to leave earlier to avoid the potential for leaving late (Figure B1a, red and blue) (Figure B1b). While the greatest fitness is gained by being in a large group, there is also a pitfall if the animals frequently change group size (e.g., in fission-fusion populations) and the animals do not know what size group they are in. Because animals in large groups are perched precariously at the objective optimal time, if those animals then move to a very small group, there could be a severe cost in fitness (Figure B1c). On the other hand, animals in small groups, when moved into a large group, suffer less of a fitness cost. Therefore, in such populations, there may be a selection pressure to leave early regardless of group size, negating the benefits of the wisdom of crowds.

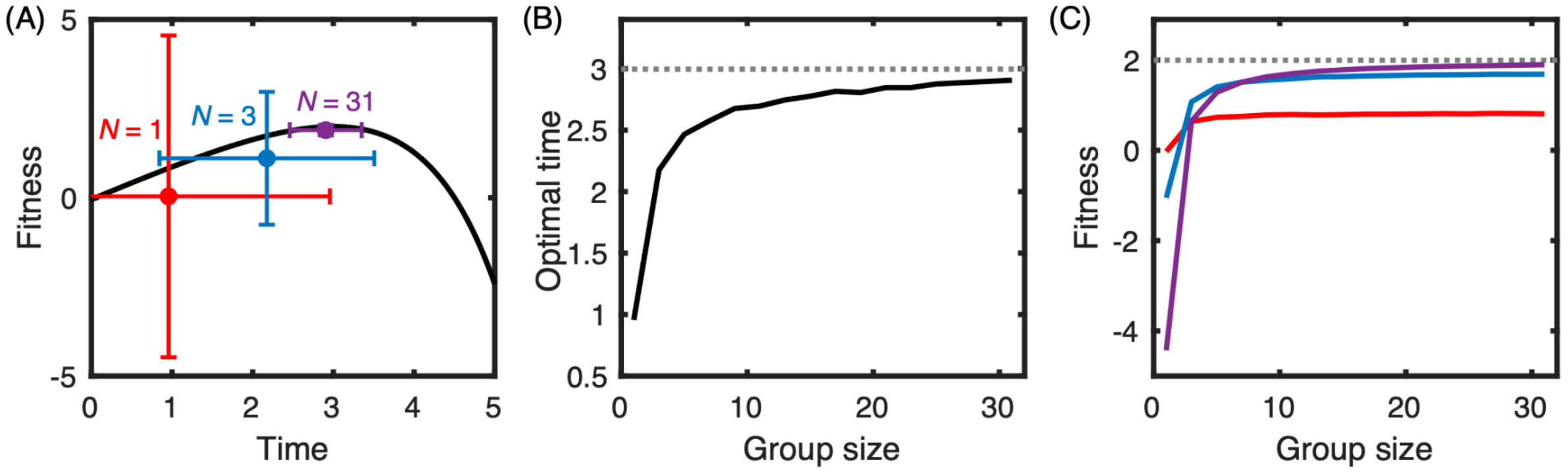

**Figure B1. Simple model of the consequences of asymmetric time costs.** (A) Fitness is asymmetric in time, where leaving early is less costly than leaving late (black curve). Smaller group sizes should leave early, but large groups can afford to be poised near the objective optimal leaving time. (B) The optimal leaving time scales nonlinearly with group size (black curve). Dotted line shows the objective optimal leaving time (i.e., the peak of the black curve in panel A). (C) The fitness when using the optimal leaving time for a particular group size (line colors, which match the colors in panel A) but transplanted to a different group size (x-axis). Using the optimal leaving time of a large group can have severe fitness costs when in a very small group (purple curve), but using the optimal leaving time of a small group has less severe costs (red and blue curves).

**Box 2. Model of the speed-accuracy trade-off in temporal decisions.**

We use a simple model to demonstrate how the speed-accuracy trade off is fundamentally altered in the time domain. For each trial we select an optimal value (location/time), $\theta_n$, by rounding the result of a normal distribution centered at $\mu_\theta$ with width $\sigma_\theta$. During each trial, at each timestep (arbitrary time units), each individual makes a new independent estimate by drawing from a normal distribution centered around the correct value, $\theta_n$, and with width $\sigma_n$ (value consistent across trials). Individuals then update their individual average estimate by taking the mean of all of their estimates up to that point. Groups update their group estimate by taking the mean of all group members' individual average estimates. For both spatial and temporal estimates, after a set number of time steps (time taken to make decision), each group selects an option by rounding their group estimate to the nearest integer. However, for temporal estimates, if the selected option is *before* the current time, their selection is updated to the current time. We calculate the error as the absolute value of the difference between the selected option (spatial or temporal) and the optimal value, $\theta_n$. Finally, we average over trials to get the mean error as a function of time taken to make the decision (Figure B2).

For spatial decisions, in which any option is available after taking any length of time to make a decision, the error decreases approximately exponentially with increasing time taken (dashed curves). This pattern highlights the classic speed-accuracy trade off. In contrast, for temporal decisions, in which options are culled as time passes, the speed accuracy trade relationship is non-monotonic (solid curves). For short decision times the temporal error drops off similarly to the spatial error. For intermediate decision times the temporal error is (slightly) less than the spatial error because in the temporal context the groups are protected against making large errors by selecting times that are too early (in the past). This effect is reduced for larger groups because outlier guesses become rarer. For longer decision times the error in temporal estimates increases dramatically (linearly) because, despite the group making an accurate decision, by the time the decision is made the opportunity to select that option has passed.

For each trial (choice of optimal day, $\theta_n$) the error is minimized when the time taken to make the decision matches $\theta_n$ (inset). However, the increase in error after $t = \theta_n$ is much steeper, so when we average over time-error curves with different values of $\theta_n$, the minimum is shifted to the left (shorter decision times). This effect is stronger for larger groups as the asymmetry of their individual trial curves is higher.

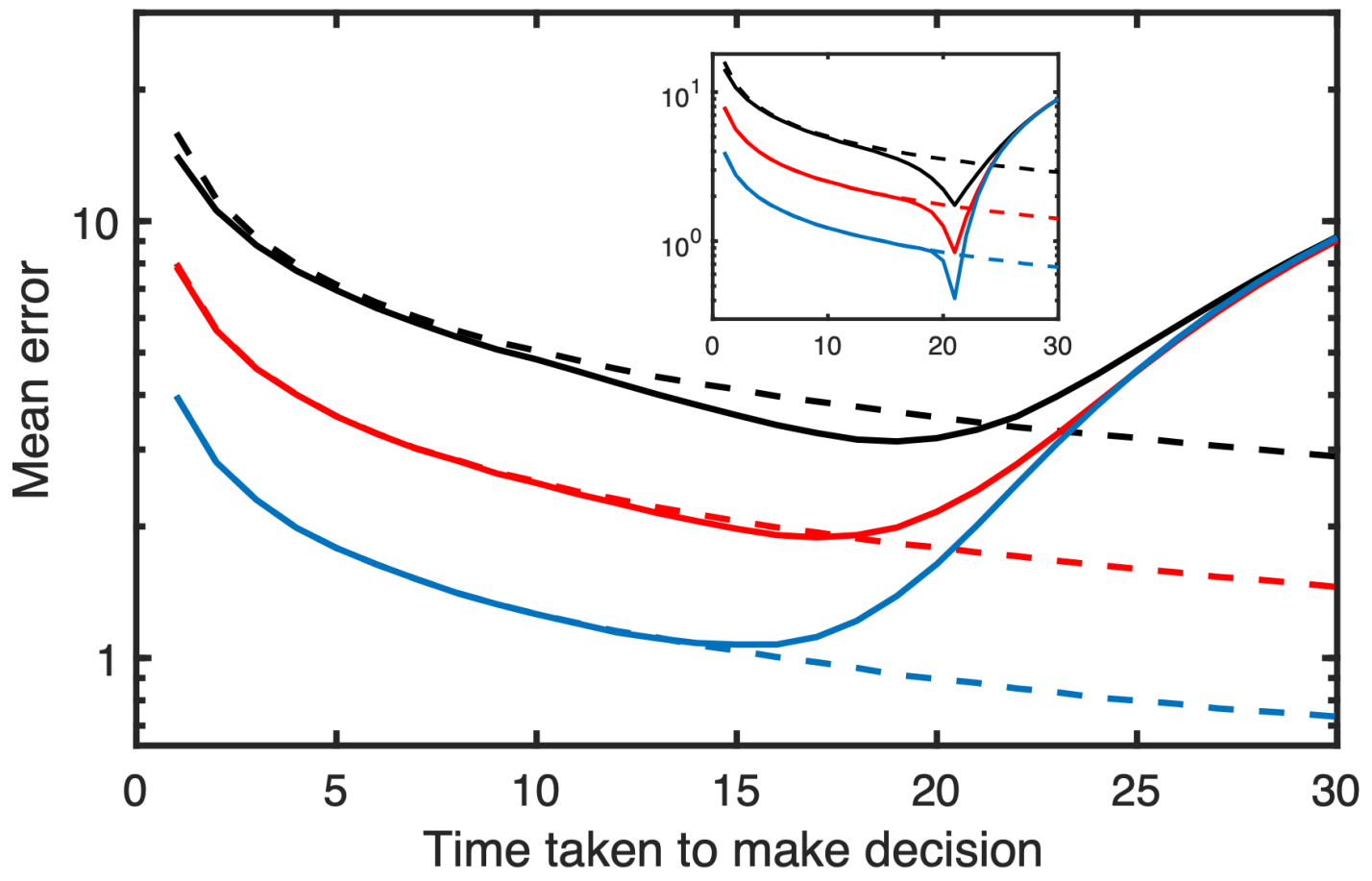

**Figure B2. Speed-accuracy trade off in temporal vs spatial decisions.** For spatial decisions (dashed lines) error drops monotonically with increased decision time. For temporal decisions (solid lines) error is minimized at a finite value of decision time and error *increases* with additional decision time after this point. The inset shows the situation in which the optimal leaving time consistently at $t = 21$ (*i.e.,* $\mu_\theta = 21$; $\sigma_\theta = 0$). In this, somewhat trivial, case it is optimal to collect information up until the optimal time. However, in the more realistic situation, in which the optimal time is not known in advance (main figure; $\mu_\theta = 21$; $\sigma_\theta = 5$), the optimal time to make a decision is less than the expected value of the optimal time and decreases with increasing group size. Colors correspond to different group sizes: $N = 1$ (black/top), $N = 4$ (red/middle), $N = 16$ (blue/bottom).

## References

1. Aikens, E. O., Bontekoe, I. D., Blumenstiel, L., Schlicksupp, A., & Flack, A. (2022). Viewing animal migration through a social lens. *Trends in Ecology & Evolution* 37:P985-996.
2. Axelrod, R., Daymude, J. J., & Forrest, S. (2021). Preventing extreme polarization of political attitudes. *Proceedings of the National Academy of Sciences* 118:e2102139118.
3. Bak-Coleman, J. B., Alfano, M., Barfuss, W., Bergstrom, C. T., Centeno, M. A., et al. (2021). Stewardship of global collective behavior. *Proceedings of the National Academy of Sciences* 118:e2025764118.
4. Bauer, S., Lisovski, S., & Hahn, S. (2016). Timing is crucial for consequences of migratory connectivity. *Oikos* 125:605-612.
5. Becker, J., Brackbill, D., & Centola, D. (2017). Network dynamics of social influence in the wisdom of crowds. *Proceedings of the National Academy of Sciences* 114: E5070-E5076.
6. Berdahl, A. M., Kao, A. B., Flack, A., Westley, P. A., Codling, E. A., et al. (2018). Collective animal navigation and migratory culture: from theoretical models to empirical evidence. *Philosophical Transactions of the Royal Society B: Biological Sciences* 373: 20170009.
7. Berdahl, A., Torney, C. J., Ioannou, C. C., Faria, J. J., & Couzin, I. D. (2013). Emergent sensing of complex environments by mobile animal groups. *Science* 339:574-576.
8. Biro, D., Freeman, R., Meade, J., Roberts, S., & Guilford, T. (2007). Pigeons combine compass and landmark guidance in familiar route navigation. *Proceedings of the National Academy of Sciences* 104:7471-7476.
9. Biro, D., Sasaki, T., & Portugal, S. J. (2016). Bringing a time–depth perspective to collective animal behaviour. *Trends in Ecology & Evolution* 31:550-562.
10. Black, J. M. (1988). Preflight signalling in swans: a mechanism for group cohesion and flock formation. *Ethology* 79:143-157.
11. Bogacz, R., Wagenmakers, E. J., Forstmann, B. U., & Nieuwenhuis, S. (2010). The neural basis of the speed–accuracy tradeoff. *Trends in Neurosciences* 33:10-16.
12. Bontekoe, I. D., Hilgartner, R., Fiedler, W., & Flack, A. (2023). The price of being late: Short- and long-term consequences of a delayed migration timing. *Proceedings of the Royal Society B: Biological Sciences* 290:20231268.
13. Centola, D., Becker, J., Zhang, J., Aysola, J., Guilbeault, D., & Khoong, E. (2023). Experimental evidence for structured information–sharing networks reducing medical errors. *Proceedings of the National Academy of Sciences* 120:e2108290120.
14. Chapman, J. W., Klaassen, R. H., Drake, V. A., Fossette, S., Hays, G. C., et al. (2011). Animal orientation strategies for movement in flows. *Current Biology* 21:R861-R870.
15. Charnov, E. L. (1976). Optimal foraging, the marginal value theorem. *Theoretical Population Biology* 9:129-136
16. Chittka, L., Skorupski, P., & Raine, N. E. (2009). Speed–accuracy tradeoffs in animal decision making. *Trends in Ecology & Evolution* 24:400-407.
17. Churchland, A., Kiani, R. & Shadlen M. (2008) Decision-making with multiple alternatives. *Nature Neuroscience* 11:693–702.

18. Cohen, J. M., Lajeunesse, M. J., & Rohr, J. R. (2018). A global synthesis of animal phenological responses to climate change. *Nature Climate Change* 8:224-228.
19. Conradt, L., & Roper, T. J. (2003). Group decision-making in animals. *Nature* 421:155-158.
20. Conradt, L., & Roper, T. J. (2005). Consensus decision making in animals. *Trends in Ecology & Evolution* 20:449-456.
21. Conradt, L., & Roper, T. J. (2007). Democracy in animals: the evolution of shared group decisions. *Proceedings of the Royal Society B: Biological Sciences* 274:2317-2326.
22. Conradt, L., & Roper, T. J. (2010). Deciding group movements: where and when to go. *Behavioural Processes* 84:675-677.
23. Cooper Jr, W. E., & Frederick, W. G. (2007). Optimal flight initiation distance. *Journal of Theoretical Biology* 244:59-67.
24. Couzin, I. D., Ioannou, C. C., Demirel, G., Gross, T., Torney, C. J., et al. (2011). Uninformed individuals promote democratic consensus in animal groups. *Science* 334:1578-1580.
25. Couzin, I. D., Krause, J., Franks, N. R., & Levin, S. A. (2005). Effective leadership and decision-making in animal groups on the move. *Nature* 433:513-516.
26. Crick, H. Q., & Sparks, T. H. (1999). Climate change related to egg-laying trends. *Nature* 399:423-423.
27. Davis, G.H., Crofoot, M.C., & Farine, D.R. (2022) Using optimal foraging theory to infer how groups make collective decisions. *Trends in Ecology & Evolution* 37:942-952.
28. Dibnah, A. J., Herbert-Read, J. E., Boogert, N. J., McIvor, G. E., Jolles, J. W., & Thornton, A. (2022). Vocally mediated consensus decisions govern mass departures from jackdaw roosts. *Current Biology* 32:R455-R456.
29. Dingle, H. (2014). *Migration: The Biology of Life on the Move*. Oxford University Press, USA.
30. Domingos, E., Grujic, J., Burguillo, J., Kirchsteiger, G., Santos, F., & Lenaerts, T. (2020). Timing Uncertainty in Collective Risk Dilemmas Encourages Group Reciprocation and Polarization. *iScience* 23.
31. Flack, A., Fiedler, W., Blas, J., Pokrovsky, I., Kaatz, M., et al. (2016). Costs of migratory decisions: a comparison across eight white stork populations. *Science Advances* 2:e1500931.
32. Freeman, P. R. (1983). The secretary problem and its extensions: A review. *International Statistical Review/Revue Internationale de Statistique* 189-206.
33. Fury, C. A., & Harrison, P. L. (2011). Seasonal variation and tidal influences on estuarine use by bottlenose dolphins (*Tursiops aduncus*). *Estuarine, Coastal and Shelf Science* 93:389-395.
34. Gall, G. E., Strandburg-Peshkin, A., Clutton-Brock, T., & Manser, M. B. (2017). As dusk falls: Collective decisions about the return to sleeping sites in meerkats. *Animal Behaviour* 132:91-99.
35. Garcia, R. A., Cabeza, M., Rahbek, C., & Araújo, M. B. (2014). Multiple dimensions of climate change and their implications for biodiversity. *Science* 344:1247579.

36. Gerlagh, R., Kverndokk, S., & Rosendahl, K. E. (2009). Optimal timing of climate change policy: Interaction between carbon taxes and innovation externalities. *Environmental and resource Economics* 43:369-390
37. Gienapp, P., & Bregnballe, T. (2012). Fitness consequences of timing of migration and breeding in cormorants. *PLoS ONE* 7:e46165.
38. Golub, B., & Jackson, M. O. (2010). Naive learning in social networks and the wisdom of crowds. *American Economic Journal: Microeconomics* 2:112-149.
39. Ha, D., & Tang, Y. (2022). Collective intelligence for deep learning: A survey of recent developments. *Collective Intelligence* 1:26339137221114874.
40. Helm, B., Piersma, T., & Van der Jeugd, H. (2006). Sociable schedules: interplay between avian seasonal and social behaviour. *Animal Behaviour* 72:245-262.
41. Hilbe, C., Chakra, M., Altrock, P., & Traulsen, A. (2013). The evolution of strategic timing in collective-risk dilemmas. *PLoS ONE* 8:e66490.
42. Horton, K. G., La Sorte, F. A., Sheldon, D., Lin, T. Y., Winner, K.,et al. (2020). Phenology of nocturnal avian migration has shifted at the continental scale. *Nature Climate Change* 10:63-68.
43. IPCC, 2022: *Climate Change 2022: Impacts, Adaptation and Vulnerability*. Contribution of Working Group II to the Sixth Assessment Report of the Intergovernmental Panel on Climate Change [H.-O. Pörtner, D.C. Roberts, M. Tignor, E.S. Poloczanska, K. Mintenbeck, A. Alegría, M. Craig, S. Langsdorf, S. Löschke, V. Möller, A. Okem, B. Rama (eds.)]. Cambridge University Press. Cambridge University Press, Cambridge, UK and New York, NY, USA, 3056 pp., doi:10.1017/9781009325844.
44. Iwasa, Y., & Levin, S. A. (1995). The timing of life history events. *Journal of Theoretical Biology* 172:33-42.
45. Kameda, T., Toyokawa, W., & Tindale, R. S. (2022). Information aggregation and collective intelligence beyond the wisdom of crowds. *Nature Reviews Psychology* 1: 345-357.
46. Kao, A. B., Berdahl, A. M., Hartnett, A. T., Lutz, M. J., Bak-Coleman, J. B., et al. (2018). Counteracting estimation bias and social influence to improve the wisdom of crowds. *Journal of The Royal Society Interface* 15:20180130.
47. Kao, A. B., & Couzin, I. D. (2014). Decision accuracy in complex environments is often maximized by small group sizes. *Proceedings of the Royal Society B: Biological Sciences* 281:20133305.
48. Kao, A. B., & Couzin, I. D. (2019). Modular structure within groups causes information loss but can improve decision accuracy. *Philosophical Transactions of the Royal Society B* 374:20180378.
49. Kao, A. B., Miller, N., Torney, C., Hartnett, A., & Couzin, I. D. (2014). Collective learning and optimal consensus decisions in social animal groups. *PLoS Computational Biology* 10:e1003762.
50. King, A. J., & Cowlishaw, G. (2007). When to use social information: the advantage of large group size in individual decision making. *Biology Letters* 3:137-139.
51. Klamser, P. P., and Romanczuk, P. (2021). Collective predator evasion: Putting the criticality hypothesis to the test. *PLoS Computational Biology* 17:e1008832.

52. Kluen, E., de Heij, M. E., & Brommer, J. E. (2011). Adjusting the timing of hatching to changing environmental conditions has fitness costs in blue tits. *Behavioral Ecology and Sociobiology* 65:2091-2103.
53. Krause, J., Ruxton, G.D., 2002. *Living in Groups*. Oxford University Press, Oxford.
54. Krause, J., Ruxton, G. D., & Krause, S. (2010). Swarm intelligence in animals and humans. *Trends in Ecology & Evolution* 25:28-34.
55. Kurvers, R. H., Nuzzolese, A. G., Russo, A., Barabucci, G., Herzog, S. M., & Trianni, V. (2023). Automating hybrid collective intelligence in open-ended medical diagnostics. *Proceedings of the National Academy of Sciences* 120:e2221473120.
56. Kurvers, R. H., Wolf, M., Naguib, M., & Krause, J. (2015). Self-organized flexible leadership promotes collective intelligence in human groups. *Royal Society Open Science* 2:150222.
57. La Sorte, F. A., Hochachka, W. M., Farnsworth, A., Sheldon, D., Fink, D., et al. (2015). Migration timing and its determinants for nocturnal migratory birds during autumn migration. *Journal of Animal Ecology* 84:1202-1212.
58. Laland, K. N. (2004). Social learning strategies. *Animal Learning & Behavior* 32:4-14.
59. Lee, C. H., & Lucas, A. (2014). Simple model for multiple-choice collective decision making. *Physical Review E* 90:052804
60. Lihoreau, M., Clarke, I. M., Buhl, J., Sumpter, D. J., & Simpson, S. J. (2016). Collective selection of food patches in Drosophila. *Journal of Experimental Biology* 219:668-675.
61. Majoris, J. E., Francisco, F. A., Burns, C. M., Brandl, S. J., Warkentin, K. M., & Buston, P. M. (2022). Paternal care regulates the timing, synchrony and success of hatching in a coral reef fish. *Proceedings of the Royal Society B* 289:20221466.
62. McCarty, J. P. (2001). Ecological consequences of recent climate change. *Conservation Biology* 15:320-331.
63. McNamara, J. M., Barta, Z., Klaassen, M., & Bauer, S. (2011). Cues and the optimal timing of activities under environmental changes. *Ecology Letters* 14:1183–1190.
64. F. Millán, M., Carranza, J., Pérez-González, J., Valencia, J., Torres-Porras, J., Seoane, J. M., De La Peña, E., Alarcos, S., Sánchez-Prieto, C. B., Castillo, L., Flores, A., & Membrillo, A. (2021). Rainfall decrease and red deer rutting behaviour: Weaker and delayed rutting activity though higher opportunity for sexual selection. *PLoS ONE* 16:e0244802.
65. Møller, A. P. (1994). Phenotype-dependent arrival time and its consequences in a migratory bird. *Behavioral Ecology and Sociobiology* 35:115-122.
66. Morimoto, J. (2019). Foraging decisions as multi-armed bandit problems: Applying reinforcement learning algorithms to foraging data. *Journal of Theoretical Biology* 467:48-56.
67. Mueller, T., O'Hara, R. B., Converse, S. J., Urbanek, R. P., & Fagan, W. F. (2013). Social learning of migratory performance. *Science* 341:999-1002.
68. Müller, F., Eikenaar, C., Crysler, Z. J., Taylor, P. D., & Schmaljohann, H. (2018). Nocturnal departure timing in songbirds facing distinct migratory challenges. *Journal of Animal Ecology* 87:1102-1115.
69. Nagy, M., Horicsányi, A., Kubinyi, E., Couzin, I. D., Vásárhelyi, G., Flack, A., & Vicsek, T. (2020). Synergistic benefits of group search in rats. *Current Biology* 30:4733-4738.

70. Namboodiri, V. M., Mihalas, S., Marton, T. M., & Hussain Shuler, M. G. (2014). A general theory of intertemporal decision-making and the perception of time. *Frontiers in Behavioral Neuroscience* 8:61.
71. Oestreich, W.K., Aiu, K.M., Crowder, L.B., McKenna, M.F., Berdahl, A.M., Abrahms, B. (2022) The influence of social cues on timing of animal migrations. *Nature Ecology & Evolution* 6:1617-1625.
72. Papadopoulou, M., Hildenbrandt, H., Sankey, D. W., Portugal, S. J., & Hemelrijk, C. K. (2022). Emergence of splits and collective turns in pigeon flocks under predation. *Royal Society Open Science* 9:211898.
73. Parker, G. A., & Smith, J. M. (1990). Optimality theory in evolutionary biology. *Nature* 348:27-33.
74. Radcliffe, K., Lyson, H. C., Barr-Walker, J., & Sarkar, U. (2019). Collective intelligence in medical decision-making: a systematic scoping review. *BMC Medical Informatics and Decision Making* 19:1-11.
75. Ratcliff, R., Smith, P. L., Brown, S. D., & McKoon, G. (2016). Diffusion decision model: Current issues and history. *Trends in Cognitive Sciences* 20:260-281.
76. Redshaw, J., & Bulley, A. (2018). Future-thinking in animals. *The Psychology of Thinking about the Future*, 31-51.
77. Ricke, K. L., & Caldeira, K. (2014). Natural climate variability and future climate policy. *Nature Climate Change* 4:333-338.
78. Sasaki, T., & Biro, D. (2017). Cumulative culture can emerge from collective intelligence in animal groups. *Nature Communications* 8:15049.
79. Sasaki, T., Granovskiy, B., Mann, R. P., Sumpter, D. J., & Pratt, S. C. (2013). Ant colonies outperform individuals when a sensory discrimination task is difficult but not when it is easy. *Proceedings of the National Academy of Sciences* 110:13769-13773.
80. Sasaki, T., & Pratt, S. C. (2011). Emergence of group rationality from irrational individuals. *Behavioral Ecology* 22:276-281.
81. Sawyer, H., LeBeau, C. W., McDonald, T. L., Xu, W., & Middleton, A. D. (2019). All routes are not created equal: an ungulate's choice of migration route can influence its survival. *Journal of Applied Ecology* 56:1860-1869.
82. Scheuerell, M. D., Zabel, R. W., & Sandford, B. P. (2009). Relating juvenile migration timing and survival to adulthood in two species of threatened Pacific salmon (*Oncorhynchus* spp.). *Journal of Applied Ecology* 46:983-990.
83. Seeley, T. D., & Visscher, P. K. (2004). Quorum sensing during nest-site selection by honeybee swarms. *Behavioral Ecology and Sociobiology* 56:594-601
84. Seeley, T. D., Visscher, P. K., Schlegel, T., Hogan, P. M., Franks, N. R., & Marshall, J. A. (2012). Stop signals provide cross inhibition in collective decision-making by honeybee swarms. *Science* 335:108-111.
85. Sergio, F., Tanferna, A., De Stephanis, R., Jiménez, L. L., Blas, J., et al. (2014). Individual improvements and selective mortality shape lifelong migratory performance. *Nature* 515:410-413.
86. Shipley, J. R., Twining, C. W., Taff, C. C., Vitousek, M. N., Flack, A., & Winkler, D. W. (2020). Birds advancing lay dates with warming springs face greater risk of chick mortality. *Proceedings of the National Academy of Sciences* 117:25590–25594.

87. Shiryaev, A. N. (2007). *Optimal Stopping Rules* (Vol. 8). Springer Science & Business Media.
88. Simons, A. M. (2004). Many wrongs: the advantage of group navigation. *Trends in Ecology & Evolution* 19:453-455.
89. Sridhar, V. H. *et al.* (2021) The geometry of decision-making in individuals and collectives. *Proceedings of the National Academy of Sciences* 118:e2102157118 .
90. Stewart, K. J., & Harcourt, A. H. (1994). Gorillas' vocalizations during rest periods: signals of impending departure? *Behaviour* 130:29-40.
91. Strandburg-Peshkin, A., Farine, D. R., Couzin, I. D., & Crofoot, M. C. (2015). Shared decision-making drives collective movement in wild baboons. *Science* 348:1358-1361.
92. Strandburg-Peshkin, A., Twomey, C. R., Bode, N. W., Kao, A. B., Katz, Y., et al. (2013). Visual sensory networks and effective information transfer in animal groups. *Current Biology* 23:R709-R711.
93. Berdahl, A., Westley, P. A., & Quinn, T. P. (2017). Social interactions shape the timing of spawning migrations in an anadromous fish. *Animal Behaviour* 126:221-229.
94. Suddendorf, T., & Corballis, M. C. (2010). Behavioural evidence for mental time travel in nonhuman animals. *Behavioural Brain Research* 215:292-298.
95. Tätte, K., Møller, A. P., & Mänd, R. (2018). Towards an integrated view of escape decisions in birds: relation between flight initiation distance and distance fled. *Animal Behaviour* 136:75-86.
96. Tomás, G. (2015). Hatching date vs laying date: what should we look at to study avian optimal timing of reproduction? *Journal of Avian Biology* 46:107-112
97. Ward, A. J., Herbert-Read, J. E., Sumpter, D. J., & Krause, J. (2011). Fast and accurate decisions through collective vigilance in fish shoals. *Proceedings of the National Academy of Sciences* 108:2312-2315.
98. Warkentin, K. M. (2011). Plasticity of hatching in amphibians: evolution, trade-offs, cues and mechanisms. *Integrative and Comparative Biology* 51:111-127.
99. Webster, M. M., Whalen, A., & Laland, K. N. (2017). Fish pool their experience to solve problems collectively. *Nature Ecology & Evolution* 1:0135.
100. Werfel, J., Petersen, K., & Nagpal, R. (2014). Designing collective behavior in a termite-inspired robot construction team. *Science*, *343*(6172), 754-758.
101. Westley, P. A., Berdahl, A. M., Torney, C. J., & Biro, D. (2018). Collective movement in ecology: from emerging technologies to conservation and management. *Philosophical Transactions of the Royal Society B: Biological Sciences*, *373*(1746), 20170004.
102. Winkler, D. W., Jørgensen, C., Both, C., Houston, A. I., McNamara, J. M., Levey, D. J., Partecke, J., Fudickar, A., Kacelnik, A., Roshier, D., & Piersma, T. (2014). Cues, strategies, and outcomes: How migrating vertebrates track environmental change. *Movement Ecology* 2:10.
103. Wolf, M., Krause, J., Carney, P. A., Bogart, A., & Kurvers, R. H. (2015). Collective intelligence meets medical decision-making: the collective outperforms the best radiologist. *PloS ONE* 10:e0134269.
104. Woolley, A. W., Chabris, C. F., Pentland, A., Hashmi, N., & Malone, T. W. (2010). Evidence for a collective intelligence factor in the performance of human groups. *Science* 330:686-688.

105. Yang, V. C., Galesic, M., McGuinness, H., & Harutyunyan, A. (2021). Dynamical system model predicts when social learners impair collective performance. *Proceedings of the National Academy of Sciences* 118:e2106292118.
106. Ylitalo, A. K., Heikkinen, J., & Kojola, I. (2021). Analysis of central place foraging behaviour of wolves using hidden Markov models. *Ethology* 127:145-157.